\documentclass[paper]{ptephy_v1} 

\usepackage{bm}

\usepackage{color}
\usepackage{ulem}

\newcommand{\nc}{\newcommand}           
\nc{\vc}[1]     {\mbox{\boldmath $#1$}} 
\nc{\wtil}      {\widetilde}            
\nc{\bras}[1]   {\langle#1|}            
\nc{\kets}[1]   {|#1\rangle}            
\nc{\bra}       {\langle}            
\nc{\ket}       {\rangle}            
\nc{\hO}        {O}           
\nc{\HO}        {\widehat{O}}   

\begin{document}

\title{Successive variational approach with the tensor-optimized antisymmetrized molecular dynamics for the $^5$He nucleus}


\author[1,2]{Takayuki Myo}
\affil[1]{General Education, Faculty of Engineering, Osaka Institute of Technology, Osaka 535-8585, Japan
\email{takayuki.myo@oit.ac.jp}
}
\affil[2]{Research Center for Nuclear Physics (RCNP), Osaka University, Ibaraki, Osaka 567-0047, Japan}

\author[3,4]{Mengjiao Lyu}
\affil[3]{College of Science, Nanjing University of Aeronautics and Astronautics, Nanjing 210016, China}
\affil[4]{Key Laboratory of Aerospace Information Materials and Physics, Ministry of Industry and Information Technology, Nanjing 210016, China}

\author[2]{Hiroshi Toki}
\author[2]{Hisashi Horiuchi}


\begin{abstract}%
We study $^5$He variationally as the first $p$-shell nucleus in the tensor-optimized antisymmetrized molecular dynamics (TOAMD) using the bare nucleon--nucleon interaction without any renormalization.
In TOAMD, the central and tensor correlation operators promote the AMD's Gaussian wave function to a sophisticated many-body state including the short-range and tensor correlations with high-momentum nucleon pairs.
We develop a successive approach by applying these operators successively with up to double correlation operators to get converging results.
We obtain satisfactory results for $^5$He, not only for the ground state but also for the excited state, and discuss explicitly the correlated Hamiltonian components in each state.
We also show the importance of the independent optimization of the correlation functions in the variation of the total energy beyond the condition assuming common correlation forms used in the Jastrow approach.
\end{abstract}

\subjectindex{D10, D11, D13}

\maketitle

\section{Introduction}
The bare nucleon--nucleon ($NN$) interaction has a strong short-range repulsion and a strong tensor force \cite{pieper01,aoki13}.
In finite nuclei, the short-range repulsion produces a short-range correlation reducing the short-range amplitudes of nucleon pairs.
The tensor force produces a tensor correlation with a strong $D$-wave transition of nucleon pairs.
The two correlations have different characters in physics, but commonly induce the high-momentum components of nucleon motion in nuclei \cite{schiavilla07}. 

The Green's function Monte Carlo (GFMC) simulation by the Argonne group has demonstrated that
they can reproduce the binding energies and low-lying energy spectra for light nuclei up to $^{12}$C with the help of three-nucleon forces \cite{carlson15}.
In GFMC, no renormalization technique is applied to the nuclear wave function and various nuclear properties are calculated directly using the resulting wave function.
One of the important results in GFMC is that the one-pion exchange contribution on the nuclear binding energy is about 80\% of the entire contribution of the two-body interaction~\cite{pieper01}.
This pion-exchange component is a dominant source of the tensor force.
At present, this numerical method requires extreme computational time to be applied to heavier nuclei.
It is by now highly desirable to develop a new method to calculate nuclear structure with large nucleon numbers by taking care of the characteristics of the $NN$ interaction.

To this end, we have introduced the tensor-optimized antisymmetrized molecular dynamics (TOAMD) \cite{myo15,myo17a,myo17b,myo17c,myo17d,myo17e,myo18,lyu18a,lyu18b,zhao19,lyu20}, 
which is an analytical variational approach directly treating the correlations induced by nuclear force.
In TOAMD, the reference wave function is taken from AMD \cite{kanada03}, which nicely describes light nuclei with effective interactions, in particular for clustering states such as the Hoyle state in $^{12}$C.
In TOAMD, we introduce two kinds of variational correlation functions of tensor- and central-operator types to treat the correlations induced by the bare $NN$ interaction. 
The correlation functions are multiplied by the AMD wave function and the resulting many-body wave functions are used to calculate the Hamiltonian matrix elements completely.
This prescription ensures the energy-variational principle for nuclei in TOAMD.

So far, TOAMD has been applied to $s$-shell nuclei within the double products of the correlation functions,
and we have successfully reproduced the properties of $s$-shell nuclei \cite{myo17a,myo17b,myo17c,myo17d}.
It is noted that in TOAMD, multiple products of the correlation functions are introduced and each correlation function in each term is independently optimized to minimize the total energy of the nucleus,
which indicates the flexibility of the correlation functions.
Owing to this property, we have shown that the binding energies in TOAMD are better than the values obtained in the Jastrow correlation method using common forms of the correlation functions for every pair \cite{myo17b,myo17c}.
We have recently applied the concept of TOAMD to the variational description of nuclear matter starting from the $NN$ interaction \cite{myo19,yamada19}.

With the success in $s$-shell nuclei and nuclear matter, it is important to apply TOAMD to $p$-shell nuclei, where the Argonne group is only the one to perform complete variational calculations without any renormalization \cite{pieper01}.
There are several theoretical studies for $p$-shell nuclei using no-core shell model and lattice simulations with a renormalization method.
In these studies, the nuclear force is constructed first in the chiral perturbation theory for various cut-off parameters for the pion-exchange interaction, with which high-momentum components are controlled \cite{machleidt10}.
The similarity renormalization group is used further to soften the interaction \cite{bogner10,hergert16}, and then no-core shell model is used to perform diagonalization of a large Hamiltonian matrix in a limited space \cite{barrett13}.
Lattice simulations are performed with the effective field theory using the chiral nuclear force \cite{epelbaum12}.
In these methods, it is essential to avoid the high-momentum components in nuclei caused by the interaction to make the convergence of solutions faster with respect to the model space.
The present TOAMD method takes quite a different approach, where we treat the bare correlations explicitly as much as possible without any truncation and discuss the properties of the obtained nuclear wave function directly.
There are other methods where the $NN$ correlations are directly treated \cite{kamada01,hiyama03,suzuki09,gandolfi14}, and in some cases they attempt calculations of the $p$-shell nucleus, $^5$He, with the bare $NN$ interaction \cite{suzuki09}.

The purpose of this paper is to show the ability of TOAMD by applying central and tensor correlation operators successively to obtain a satisfactory result for $p$-shell nuclei.
As a first trial of TOAMD to $p$-shell nuclei, we calculate $^5$He as a five-body problem.
We investigate the structure of this nucleus within the double products of the correlation functions.
The role of each correlation function is analysed focusing on their independent optimization.
We also extend TOAMD by superposing the different reference AMD wave functions in a generator coordinate method,
which enables us to obtain the excited states as well as the ground state simultaneously.
We discuss the structures of ground and excited states of $^5$He in a unified TOAMD framework.

In TOAMD, the increase of the variational accuracy is straightforward by successively adding the higher orders of the correlation functions in the wave function.
The necessary matrix elements at any order of TOAMD are calculated in the analytical form. In general, an increase of the order of TOAMD requires more computer resources.

In Sect.~\ref{sec:model}, we explain the nuclear models of AMD and TOAMD.
In Sect.~\ref{sec:results}, we present the results of the $s$-shell nuclei $^3$H and $^4$He, and the $p$-shell nucleus $^5$He in TOAMD.
A summary is given in Sect.~\ref{sec:summary}.

\section{Tensor-optimized antisymmetrized molecular dynamics (TOAMD)}\label{sec:model}

We explain the formulation of TOAMD \cite{myo15,myo17d}. 
The reference AMD wave function $\Phi_{\rm AMD}$ is the Slater determinant of nucleons with mass number $A$ as
\begin{eqnarray}
\Phi_{\rm AMD}
&=& \frac{1}{\sqrt{A!}}\, {\rm det} \left\{ \prod_{i=1}^A \psi_{\sigma_i\tau_i}(\vc{r}_i) \right\}~,
\label{eq:AMD}
\\
\psi_{\sigma\tau}(\vc{r})&=&\left(\frac{2\nu}{\pi}\right)^{3/4} e^{-\nu(\bm{r}-\bm{D})^2} \chi_{\sigma} \chi_{\tau}.
\label{eq:Gauss}
\end{eqnarray}
The nucleon wave function $\psi_{\sigma\tau}(\vc r)$ has a Gaussian wave packet with a common range parameter $\nu$ and a centroid position $\vc D$, 
a spin part $\chi_{\sigma}$, and an isospin part $\chi_{\tau}$.
In this study, $\chi_{\sigma}$ is the up or down component and $\chi_{\tau}$ is a proton or neutron.
The AMD wave function $\Phi_{\rm AMD}$ has a set of $\vc D=\{\vc D_i\}$ with $i=1,\ldots,A$
with the condition of $\sum_{i=1}^A \vc D_i=\vc 0$.

We introduce two kinds of the two-body correlation functions, $F_D$ for the tensor force and $F_S$ for the short-range repulsion,
to make the correlated wave function. These functions are defined as
\begin{eqnarray}
F_D
&=& \sum_{t=0}^1\sum_{i<j}^A f^{t}_{D}(r_{ij})\, S_{12}(\bm{\hat{r}}_{ij})\, (\vc \tau_i\cdot \vc \tau_j)^t \,,
\label{eq:Fd}
\\
F_S
&=& \sum_{t=0}^1\sum_{s=0}^1\sum_{i<j}^A f^{t,s}_{S}(r_{ij})\,(\vc \tau_i\cdot \vc \tau_j)^t\, (\vc\sigma_i\cdot \vc\sigma_j)^s\,,
\label{eq:Fs}
\end{eqnarray}
with a relative coordinate $\vc r_{ij}=\vc r_i - \vc r_j$.
It is noted that $F_D$ and $F_S$ are scalar functions.
The pair functions $f^{t}_{D}(r)$ and $f^{t,s}_{S}(r)$ are variationally determined, as explained later. 
The labels $s$ and $t$ are the indices to express the spin--isospin dependence of the correlation functions.
The function $F_D$ in Eq.~(\ref{eq:Fd}) produces the $D$-wave transition due to the tensor operator $S_{12}$,
and $F_S$ describes the central correlation including the short-range one.
The functions $F_D$ and $F_S$ change the relative wave function of two nucleons in the AMD wave function $\Phi_{\rm AMD}$
and can excite two nucleons to high-momentum states in nuclei.

We multiply these correlation functions to $\Phi_{\rm AMD}$ and superpose these components.
We define the TOAMD wave function with the single correlation functions as
\begin{eqnarray}
\Phi_{\rm TOAMD}^{\rm single}
&=& (1+F_S+F_D) \times\Phi_{\rm AMD}\,.
\label{eq:TOAMD1}
\end{eqnarray}
We can increase the order of the correlation function in TOAMD by further adding the double products of $F_D$ and $F_S$ as
\begin{eqnarray}
\Phi_{\rm TOAMD}
&=& (1+F_S+F_D+F_S F_S+F_D F_S+F_D F_D)
\nonumber\\
&\times&\Phi_{\rm AMD}~.
\label{eq:TOAMD2}
\end{eqnarray}
In the present study, we use this form of TOAMD for calculations of nuclei up to the second orders, which are based on the power series expansion in terms of correlations $F_D$ and $F_S$ independently. 
It is noted that all of $F_D$ and $F_S$ in each term in Eq.\,(\ref{eq:TOAMD2}) are independent and variationally determined,
which means that there are four kinds of $F_D$ and four kinds of $F_S$, while we use the common notations of $F_D$ and $F_S$ in the equations for simplicity. 
This property of TOAMD brings a flexibility of the correlation functions in comparison with the so-called Jastrow method, in which common correlation functions are assumed in all nucleon pairs \cite{jastrow55}.
As an extension of Eq.~(\ref{eq:TOAMD2}), we can successively increase the order of power expansion to triple products such as $F_D F_D F_S$ when we increase the variational accuracy of the solutions.
This extension is feasible and would require more computational resources for numerical calculation in future.
The form of wave function in TOAMD given in Eq.\,(\ref{eq:TOAMD2}) is general and commonly used for all nuclei.

We use the Hamiltonian with a two-body bare $NN$ interaction $V$ for mass number $A$ as
\begin{eqnarray}
    H
&=& \sum_i^{A} t_i - T_{\rm c.m.} + \sum_{i<j}^{A} v_{ij}\, ,
    \label{eq:Ham}
\end{eqnarray}
Here, $t_i$ and $T_{\rm c.m.}$ are the kinetic energies of each nucleon and the center of mass, respectively.
In this study, we employ the AV6$^\prime$ potential \cite{wiringa95,wiringa02,wiringa_web} as $v_{ij}$ consisting of central and tensor terms
without $LS$ and Coulomb terms.
We directly use this Hamiltonian in TOAMD without any renormalization, 
which enables us to investigate the explicit roles of the short-range and tensor correlations in nuclei
on any observables such as the nucleon momentum distribution \cite{lyu20}.

\begin{figure}[t]
\centering
\includegraphics[width=7.5cm,clip]{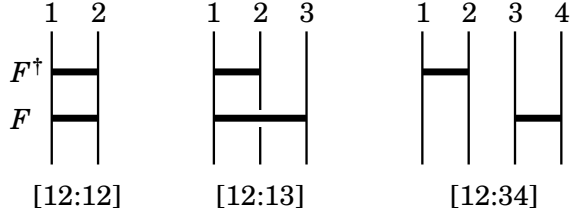}
\caption{Diagrams of the cluster expansion of $F^\dagger F$ from two-body to four-body terms,
  where the vertical and horizontal lines represent the particles and the correlation function $F$, respectively. The numbers in the square brackets describe the particle index specifying the configuration of each diagram \cite{myo17c}.}
\label{fig:FF}
\end{figure}

\begin{figure}[bt]
\centering
\includegraphics[width=15.0cm,clip]{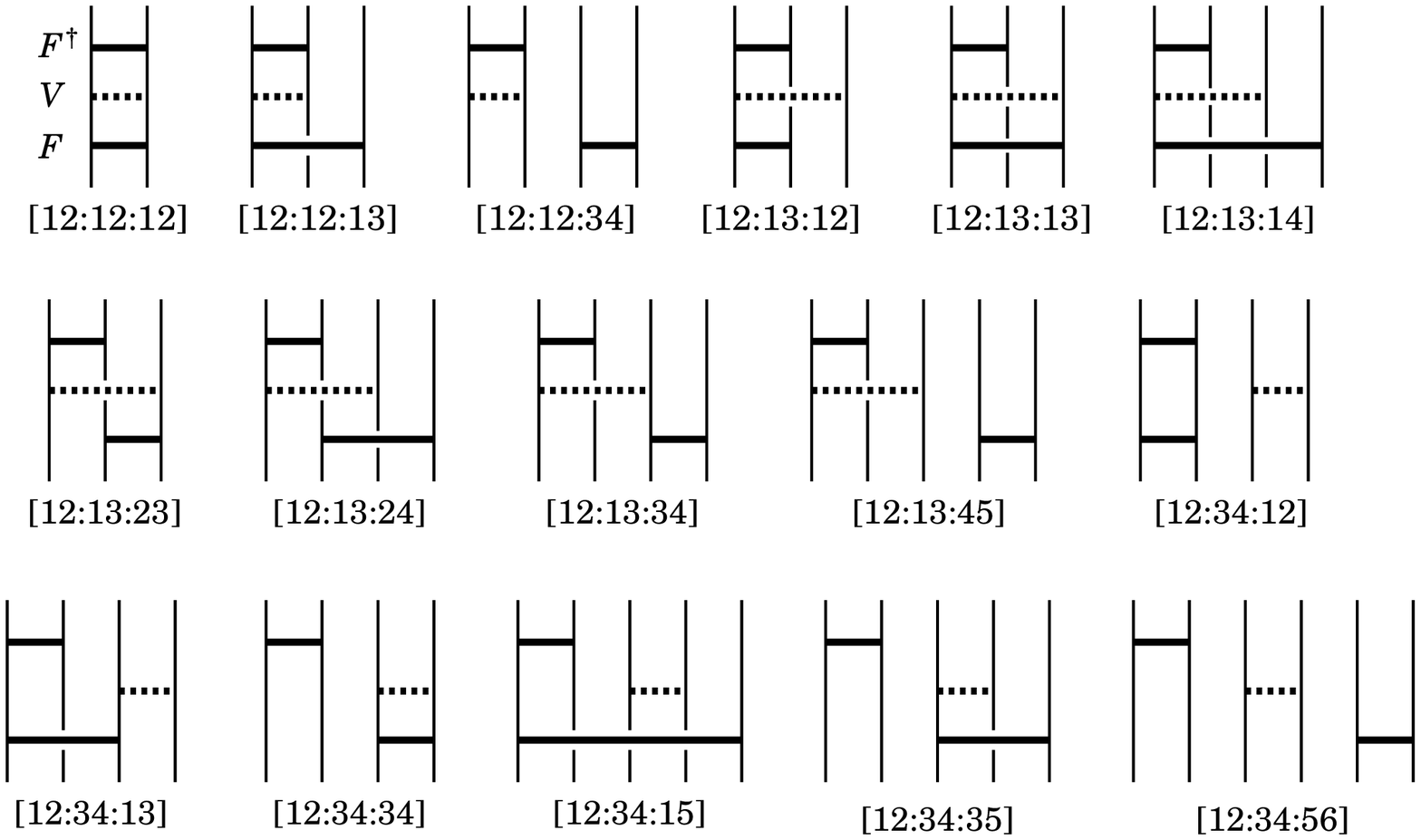}
\caption{Diagrams of the cluster expansion of $F^\dagger VF$ from two-body to six-body terms,
where $V$ represents a two-body operator with dotted lines.}
\label{fig:FFF}
\end{figure}

\begin{figure}[th]
\centering
\includegraphics[width=11.0cm,clip]{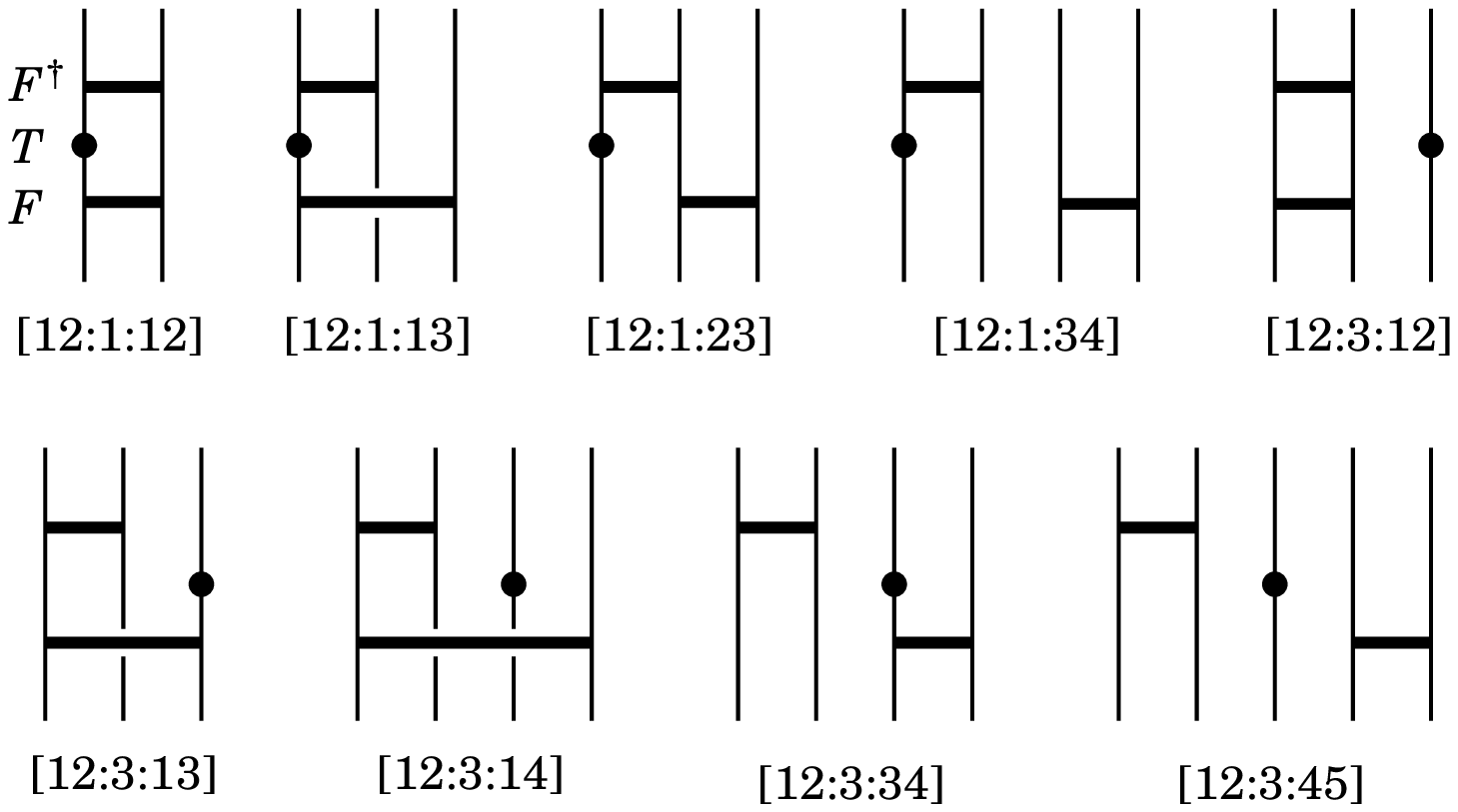}
\caption{Diagrams of the cluster expansion of $F^\dagger TF$ from two-body to five-body terms,
where $T$ represents a one-body operator with solid circles.}
\label{fig:FGF}
\end{figure}

The total energy $E$ of a nucleus in TOAMD is given as:
\begin{eqnarray}
    E
&=&\frac{\langle\Phi_{\rm TOAMD} |H|\Phi_{\rm TOAMD}\rangle}{\langle\Phi_{\rm TOAMD} |\Phi_{\rm TOAMD}\rangle}
    \nonumber
    \\
&=&\frac{\langle\Phi_{\rm AMD} | \widetilde{H} |\Phi_{\rm AMD}\rangle}{\langle\Phi_{\rm AMD} | \widetilde{N} |\Phi_{\rm AMD}\rangle}.
\label{eq:E_TOAMD}
\end{eqnarray}
We expand the TOAMD wave function $\Phi_{\rm TOAMD}$ using Eq.~(\ref{eq:TOAMD2}) and
introduce the correlated operators with tildes in Eq.~(\ref{eq:E_TOAMD}) with respect to the AMD wave function $\Phi_{\rm AMD}$. 
The operators $\widetilde{H}$ and $\widetilde{N}$ are the correlated Hamiltonian and norm operator, respectively.
They involve the multiple products of operators such as $F^\dagger H F$ and $F^\dagger F$, where $F$ stands for $F_D$ and $F_S$. 
We evaluate the matrix elements of the correlated operators with the AMD wave function.
The operators $\widetilde{H}$ and $\widetilde{N}$ consist of various products of correlation functions
and they are individually expanded into many-body operators using the cluster expansion technique \cite{myo15,myo17d}.
In the case of $F^\dagger F$, this product is expanded into two-body, three-body and four-body operators as shown in the diagrams of Fig.~\ref{fig:FF}.
For the two-body interaction $V$, the correlated interaction $F^\dagger V F$ gives up to six-body operators and their diagrams are shown in Fig.~\ref{fig:FFF}.
For one-body operators such as kinetic energy and square radius, the correlated operators give up to five-body ones and their diagrams are shown in Fig.~\ref{fig:FGF}.
Similarly, the correlated operators $F^\dagger F^\dagger T F F$ and $F^\dagger F^\dagger V F F$ give up to nine-body and ten-body operators, respectively.

In TOAMD, we adopt all of the resulting many-body operators in the cluster expansion of the correlated operators and calculate their many-body matrix elements using the AMD wave function.
This treatment is important to keep TOAMD as a variational framework.
The calculation of the matrix elements of many-body operators is performed analytically for any order of the multiple products of the correlation functions \cite{myo15,myo17d}.
In general, higher-body operators in the cluster expansion tend to require a computational cost to calculate their matrix elements.

The TOAMD wave function has two kinds of variational functions, the AMD wave function in Eq. (\ref{eq:AMD}) and the correlation functions $F$ in Eqs. (\ref{eq:Fd}) and (\ref{eq:Fs}).
We determine these functions under the Ritz variational principle for the total energy as $\delta E=0$ in Eq.~(\ref{eq:E_TOAMD}).
In the determination of the radial distribution of $F_D$ and $F_S$,
we use the Gaussian expansion to express the pair functions $f^{t}_{D}(r)$ and $f^{t,s}_{S}(r)$:
\begin{eqnarray}
   f^t_D(r)
&=&\sum_{n=1}^{N_{\rm G}} C^t_n \, e^{-a^t_n r^2},
   \label{eq:cr1}
   \\
   f^{t,s}_S(r)
&=&\sum_{n=1}^{N_{\rm G}} C^{t,s}_n\, e^{-a^{t,s}_n r^2},
   \label{eq:cr2}
\end{eqnarray}
where $a^t_n$, $a^{t,s}_n$, $C^t_n$, and $C^{t,s}_n$ are variational parameters with the index $n$. 
We take the number of Gaussian functions $N_{\rm G}=6$ to get converging solutions.
For the Gaussian ranges $a^t_n$, $a^{t,s}_n$, we search for their optimized values in a wide range.
The coefficients $C^t_n$ and $C^{t,s}_n$ are linear parameters and are determined by diagonalizing the Hamiltonian matrix.
Typical distributions of $f^t_D(r)$ and $f^{t,s}_S(r)$ are shown for $s$-shell nuclei in Ref. \cite{myo17c}.

For the double products of correlation functions such as $F_D F_D$, the products of two Gaussian functions in Eq. (\ref{eq:cr1}) are treated as single basis functions with an amplitude of $C_n^t C_{n'}^{t'}$.
In the same manner, the products of $C^t_n$ and $C^{t,s}_n$,  $C_n^t C_{n'}^{t'}$, $C_n^t C_{n'}^{t',s}$, and $C_n^{t,s} C_{n'}^{t',s'}$ become linear parameters.

Finally, we rewrite the TOAMD wave function in Eq.~(\ref{eq:TOAMD2}) in a linear combination form using the coefficients $\widetilde{C}_\alpha$ of the Gaussian expansion of the correlation functions:
\begin{eqnarray}
   \Phi_{\rm TOAMD}
&=& \sum_{\alpha=0} \widetilde{C}_\alpha\,  \Phi_{{\rm TOAMD},\alpha} \,,
   \label{eq:linear}
   \\
   H_{\alpha \beta}
&=& \langle\Phi_{{\rm TOAMD},\alpha} |H|\Phi_{{\rm TOAMD},\beta}\rangle\, ,
   \nonumber\\
   N_{\alpha \beta}
&=& \langle\Phi_{{\rm TOAMD},\alpha} |\Phi_{{\rm TOAMD},\beta}\rangle \, ,
   \nonumber
\end{eqnarray}
where the labels $\alpha$ and $\beta$ are the set of the Gaussian index $n$ and the labels $s$ and $t$ in the correlation functions,
and the summation includes all the single and double correlated states.
The case with labels of $\alpha=\beta=0$ indicates the AMD wave function.
The Hamiltonian and norm matrix are $H_{\alpha \beta}$ and $N_{\alpha \beta}$, respectively.
We solve the generalized eigenvalue problem to determine the total energy $E$ and the coefficients $\widetilde{C}_\alpha$ in Eq.~(\ref{eq:linear}):
\begin{eqnarray}
   \sum_{\beta=0} \left( H_{\alpha\beta} - E\, N_{\alpha\beta} \right) \widetilde{C}_\beta &=&0.
   \label{eq:eigen}
\end{eqnarray}

We explain the procedure to evaluate the matrix elements of the correlated Hamiltonian and norm using the AMD wave function in Eq.~(\ref{eq:E_TOAMD}).
We express the $NN$ interaction $V$ as a sum of Gaussians, similar to the correlation functions.
In the cluster expansion of $\widetilde{H}$ and $\widetilde{N}$, many-body operators have various combinations of the square of the interparticle coordinates $\vc r_{ij}^{\,2}$ in the Gaussians. 
We perform a Fourier transformation of each Gaussian with the individual momentum $\vc k$, which results in the product of the plane waves, $e^{i\bm{k}\cdot \bm{r}_i}\, e^{-i\bm{k}\cdot \bm{r}_j}$.
We calculate the single-particle matrix elements of the plane waves with various momenta in AMD using Eq.~(\ref{eq:Gauss}).
Using these matrix elements, we perform multiple integration over all momenta and obtain the correlated matrix elements.
Typical analytical expressions of the many-body matrix elements are given in Refs. \cite{myo15,myo17d}.

In the present study, we adopt the intrinsic AMD wave function $\Phi_{\rm AMD}$ in Eq.~(\ref{eq:AMD}),
which is a mixed state of the $J^\pi$ components.
In TOAMD, the fraction of the $J^\pi$ components in AMD is changeable in the total wave function given in Eq.~(\ref{eq:TOAMD2})
due to the degrees of freedom of the correlation functions $F$.
The function $F$ is expanded in a linear combination form using the Gaussian functions in Eqs.~(\ref{eq:cr1}) and (\ref{eq:cr2}), and the expansion coefficients are optimized by solving the energy eigenvalue problem in Eq.~(\ref{eq:eigen})
and depend on the eigenstates of TOAMD.
In general, the function $F$ works to ensure that the TOAMD wave function has good quantum numbers.
We have checked this property in $^3$H and $^4$He using the non $s$-wave configurations of $\Phi_{\rm AMD}$,
and the results of TOAMD within $F^2$ show their ground states with positive parity.

We can extend TOAMD superposing the various AMD wave functions in the generator coordinate method (GCM).
We express the different AMD wave functions $\Phi_{{\rm AMD},k}$ with the index $k$,
in which the set of Gaussian centroid positions $\vc D$ is different in Eq.~(\ref{eq:AMD}).
The number of the AMD wave functions is $N_{\rm GCM}$ .
In TOAMD, the correlation functions also depend on the index $k$.
We call this method TOAMD+GCM and the corresponding total wave function $\Phi_{\rm GCM}$ is given as
\begin{eqnarray}
\Phi_{\rm GCM}
&=& \sum_{k=1}^{N_{\rm GCM}} \Phi_{{\rm TOAMD},k},
\label{eq:GCM}
\\
\Phi_{{\rm TOAMD},k}
&=& (1+F_{S,k}+F_{D,k}+F_{S,k}\, F_{S,k}
\nonumber\\
&+& F_{D,k}\, F_{S,k} + F_{D,k}\, F_{D,k})\times \Phi_{{\rm AMD},k}
\nonumber\\
&=& \sum_\alpha \widetilde{C}_{\alpha,k}\, \Phi_{{\rm TOAMD},\alpha,k}\, .
\end{eqnarray}
The amplitudes $\{\widetilde{C}_{\alpha,k}\}$ are determined in the minimization of the total energy $E_{\rm GCM}$ as an eigenvalue problem in the same form as Eq. (\ref{eq:eigen}).

\section{Results}\label{sec:results}

\subsection{$s$-shell nuclei}
We start from the single configuration of AMD in TOAMD in Eq. (\ref{eq:TOAMD2}).
We discuss the applicability of TOAMD showing the results of $s$-shell nuclei, $^3$H and $^4$He,
in which $s$-wave configurations of the AMD wave function are used with the centroid parameters ${\vc D}={\vc 0}$ for all nucleons.
This condition has been shown to be variationally favored \cite{myo17a}.
In Table.~\ref{tab:ene_3H_4He}, we summarize the energies of two nuclei with two kinds of the range parameter $\nu$ in the Gaussian wave packet in Eq.~(\ref{eq:Gauss}); 
One is $\nu = 0.15$ fm$^{-2}$, which is adopted to minimize the total energy of $^5$He.
Another choice is that $\nu$ is optimized for each nucleus.
It is found that the energy differences between the two kinds of $\nu$ values are small.
This means that the solutions are less dependent on $\nu$ in TOAMD.
This is because the correlation functions can optimize the TOAMD wave function to minimize the total energy.
The GFMC calculation provides $-7.95$ MeV for $^3$H and $-26.8$5 MeV for $^4$He after subtracting the Coulomb force contribution, which amounts to $0.7$ MeV repulsion for $^4$He \cite{gandolfi14}.
We see that the energies obtained in TOAMD are close to these values in two nuclei as shown in Table \ref{tab:ene_3H_4He}.
From these comparisons, our TOAMD solutions provide sufficiently reliable energies for $s$-shell nuclei, and we are able to discuss the spectroscopic properties of each nucleus. 

When we want to increase the variational accuracy to be close to rigorous calculations such as GFMC,
more correlation terms such as the triple products of $F^3$, would be successively added, although more computing power would be demanded.

\begin{table}[t]
\begin{center}
\caption{Total energies $E$ of $^3$H ($\frac12^+$) and $^4$He ($0^+$) in TOAMD with the AV6$^\prime$ potential in units of MeV.
Two kinds of $\nu$ values in the Gaussian are used. One is $\nu=0.15$ fm$^{-2}$ optimized for $^5$He and the other is the optimized $\nu$ for each nucleus, denoted as ``Optimized $\nu$''.}
\label{tab:ene_3H_4He}
\begin{tabular}{c|cccc}
\noalign{\hrule height 0.5pt}
        &~~$\nu=0.15$~~~&~~~Optimized $\nu$~~~&~~~GFMC \cite{wiringa02,wiringa_web,gandolfi14} \\
\noalign{\hrule height 0.5pt}
$^3$H   &  ~$-7.89$    &  ~$-7.93$ ($\nu=0.11$) & ~$-7.95(1)$ \\
$^4$He  &  $-25.39$    &  $-26.01$ ($\nu=0.23$) & $-26.85(2)$ \\
\noalign{\hrule height 0.5pt}
\end{tabular}
\end{center}
\end{table}

\begin{table}[t]
\begin{center}
  \caption{Total energies $E$ of the ground state of $^5$He in TOAMD with AV6$^\prime$ potential in units of MeV.
  We add the correlation terms successively.}
\label{tab:TOAMD_E}
\begin{tabular}{c|cccccc}
\noalign{\hrule height 0.5pt}
~~AMD~~  & ~~$+S$~~  &~~$+D$~~   &~~$+SS$~~&~~$+DS$~~ &~~$+DD$~~  \\
\noalign{\hrule height 0.5pt}
$45.72$  &  $~10.92$  & $-9.96$  & $-13.58$ & $-18.16$ & $-20.33$       \\
\noalign{\hrule height 0.5pt}
\end{tabular}
\end{center}
\end{table}

\begin{figure}[t]
\centering
\includegraphics[width=9.0cm,clip]{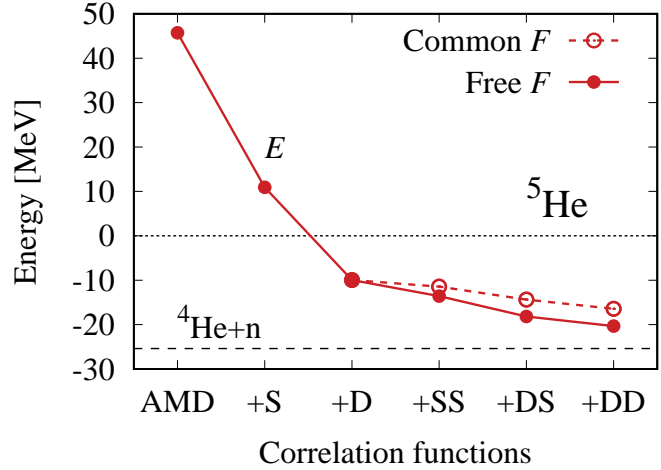}
\caption{Total energy $E$ of the ground state of $^5$He with the AV6$^\prime$ potential by adding each correlation term successively. The solid circles are the results with full variation of the correlation functions $F$.
The open circles are the results assuming the common correlation functions. The dashed horizontal line indicates the $^4$He+$n$ threshold energy.}
\label{fig:ene_5He_E}
\end{figure}

\begin{figure}[t]
\centering
\includegraphics[width=9.0cm,clip]{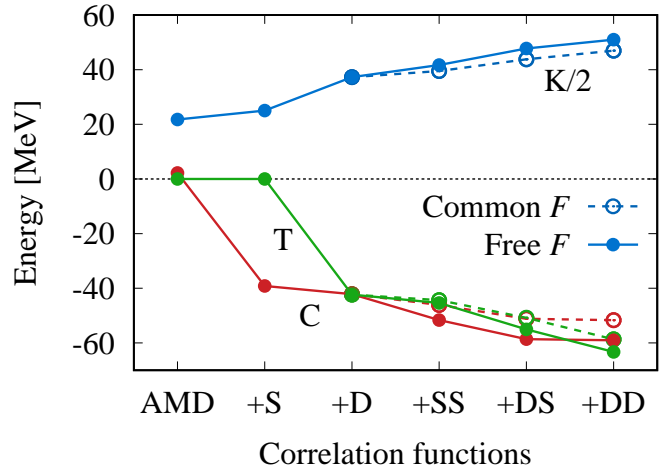}
\caption{Components of the kinetic energy $K$, central force $C$, and tensor force $T$ in the ground state of $^5$He with the AV6$^\prime$ potential.
The half value of kinetic energy is shown denoted as $K/2$.
The solid circles are the results with full variation of the correlation functions $F$.
The open circles are the results assuming the common correlation functions.}
\label{fig:ene_5He_H}
\end{figure}

\subsection{$^5$He with single configuration of AMD}
We calculate $^5$He in TOAMD with $\nu=0.15$ fm$^{-2}$ and start from the single $^4$He+$n$ cluster configuration of the AMD wave function with an $s$-wave state of $^4$He, namely, $\bm{D}_i=(0,0,-d/5)$ for $i=1,\ldots,4$ and $\bm{D}_5=(0,0,4d/5)$
for the Gaussian centroid parameters with a cluster distance $d$.
It is noted that the $^4$He nucleus in $^5$He is not the $s$-wave state in TOAMD due to the correlation functions $F$,
which induce the excitation of nucleons from the $s$-wave state.

We take the cluster distance $d$ as $0.2$ fm, which is determined variationally under the bound state approximation.
In this configuration, the last neutron occupies the $p$-orbit due to the antisymmetrization
and the parity is obtained as $-1.00$ in the AMD wave function $\Phi_{\rm AMD}$. 
In Table \ref{tab:TOAMD_E}, the total energy of the $^5$He ground state is shown by adding the correlation terms successively,
which corresponds to extending the variational space.
The same results are shown using the solid line in Fig.~\ref{fig:ene_5He_E}.
We start from the case of the AMD wave function, where the resulting energy is very high as denoted by AMD in Fig. \ref{fig:ene_5He_E}.
We then add the correlation functions step by step and see the converging behavior in energy.
The label of $+$S means the addition of the $F_S$ component and the TOAMD wave function is $(1+F_S)\, \Phi_{\rm AMD}$.
The $+$DD case is the full component of the TOAMD wave function defined in Eq.~(\ref{eq:TOAMD2}).
The final energy is $-20.3$ MeV and is located above the $^4$He+$n$ threshold energy by about 5.7 MeV
because of the unbound nature of the $^5$He system.

In this study, we do not perform parity projection on the TOAMD wave function, since the final state has an expectation value of parity of $-1.00$.
Hence, the resulting state has odd parity, and we can identify it with the $p$-wave state of $^5$He. 
According to the GFMC results with AV6$^\prime$ \cite{wiringa02,wiringa_web}, the energy of $^5$He
is $-24.23(4)$ MeV for the $3/2^-$ state and $-24.15(3)$ MeV for the $1/2^-$ state. The $p$-orbit splitting is small at around 80 keV. 
From this fact and also considering the optimization of the quantum number in TOAMD by the correlation functions $F$ explained in Sect.~\ref{sec:model},
we estimate that the effect of angular momentum projection is small in the present calculation of TOAMD.
Among the $3/2^-$ and $1/2^-$ components, we consider that the resulting ground state of $^5$He can be dominated by the $3/2^-$ state, which is favored in energy.

In Fig.~\ref{fig:ene_5He_H}, the Hamiltonian components of the $^5$He ground state are shown
by adding correlation terms in a similar way to that shown in Fig.~\ref{fig:ene_5He_E}. It is found that as correlation terms are added
successively, every component increases its magnitude and more correlations are involved in the wave function.
In particular, the enhancement of the kinetic energy indicates the inclusion of the high-momentum component
in the wave function induced by short-range repulsion and tensor force by the correlation functions.
In TOAMD, we can discuss explicitly the contribution of each Hamiltonian component at each step of the variational space 
without any renormalization, which is the advantage of TOAMD.

In TOAMD, the correlation functions $F_S$ and $F_D$ are optimized independently in each term of Eq.~(\ref{eq:TOAMD2})
at each variational space.
It is interesting to see this effect on the solutions of $^5$He, and for this purpose we perform the following calculation.
First, $F_S$ and $F_D$ are determined in the single correlation function of TOAMD as $\Phi^{\rm single}_{\rm TOAMD}$, defined in Eq.~(\ref{eq:TOAMD1}).
Second, keeping the radial form of $F_S$ and $F_D$ with Gaussian expansion in all correlation terms, 
we perform the calculation including double correlation functions,
where the weights of the double correlation functions are variational parameters.
This calculation corresponds to the Jastrow correlation method
in which every nucleon pair in nuclei is correlated by the common correlation function.
Under this condition, TOAMD provides the energies of $^5$He as $-16.4$ MeV.
The energy loss from the full calculation is about 4 MeV, which is larger than the 2 MeV of $^4$He \cite{myo17c}.
This amount indicates the importance of the independent optimization of the correlation functions in TOAMD
and this property contributes to the good energy convergence.
We show this effect step by step at each correlation level using open circles connected by dashed lines in Figs.~\ref{fig:ene_5He_E} and \ref{fig:ene_5He_H}.
In the full TOAMD calculation with up to the $F_DF_D$ term,
each Hamiltonian component show clear differences by more than 5 MeV. 
Hence, the variation of both correlation functions of $F_S$ and $F_D$ contributes to the optimization of the total wave function.

\subsection{$^5$He with the generator coordinate method (GCM)}
We extend TOAMD by superposing the TOAMD basis states having different AMD wave functions
according to Eq.~(\ref{eq:GCM}) and investigate this effect on $^5$He.
In TOAMD+GCM, we can discuss the structures of not only the ground state but also the excited state of $^5$He.
In this study, we make three basis states with the distances $d$ between $^4$He and $n$ being 0.2 fm, 0.5 fm, and 1.0 fm.
This choice is sufficient to include the effect of the valence-neutron motion of $^5$He without continuum states.
It is noted that, when we start from the different AMD configurations of $^3$H+$d$ with the $s$-wave configurations for both $^3$H and $d$,
the resulting energies are almost the same as those obtained in the $^4$He+$n$ case for both the ground and excited states.
This fact indicates that the correlation functions in TOAMD can optimize the total wave function
even starting from different reference AMD wave functions.

\begin{figure}[t]
\centering
\includegraphics[width=9.0cm,clip]{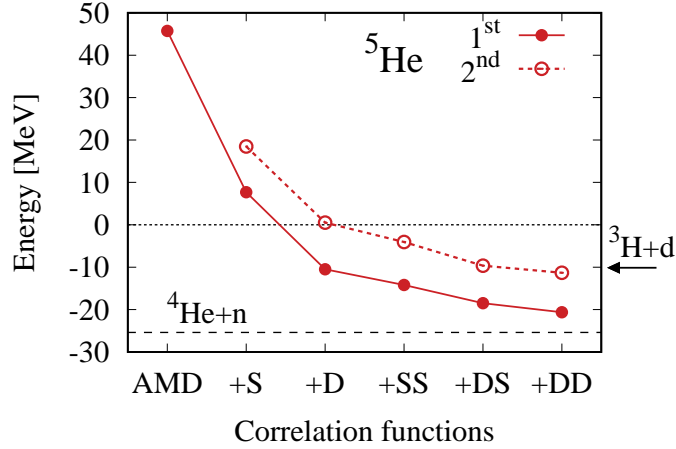}
\caption{Total energy $E_{\rm GCM}$ of the ground (1$^{\rm st}$) and excited (2$^{\rm nd}$) states of $^5$He with the AV6$^\prime$ potential by adding each correlation term successively.
The solid (open) circles indicate the ground (excited) state. The horizontal arrow on the right-hand side indicates the theoretical threshold energy of the $^3$H+$d$ state.}
\label{fig:ene_5He_E2}
\end{figure}

\begin{table}[t]
\begin{center}
\caption{Total energies $E_{\rm GCM}$ of the ground and excited states of $^5$He in TOAMD+GCM with the AV6$^\prime$ potential in units of MeV.}
\label{tab:GCM_E}
\begin{tabular}{c|cccccc}
\noalign{\hrule height 0.5pt}
     & ~~$+S$~~   & ~~$+D$~~  &~~$+SS$~~ & ~~$+DS$~~ &~~$+DD$~~  \\
\noalign{\hrule height 0.5pt}
ground state ~ &  $~7.68$  & $-10.50$  & $-14.19$ & $-18.49$ & $-20.63$       \\
excited state~ &  $18.47$  & $~~0.52$  & $\;-4.07$& $\;-9.65$ & $-11.32$       \\
\noalign{\hrule height 0.5pt}
\end{tabular}
\end{center}
\end{table}

We show the results of $^5$He in TOAMD+GCM in Table~\ref{tab:GCM_E}, 
in which the successive addition of correlation terms is performed. The results are also shown in Fig.~\ref{fig:ene_5He_E2}.
For the ground state, its energy is obtained as $-20.63$ MeV, which shows an energy gain of 0.4 MeV from the value of TOAMD
with a single AMD wave function.
This result indicates that most of the important correlations are described by the correlation functions in TOAMD
and the GCM effect on the ground state energy is not large.
In Table~\ref{tab:GCM_H}, we show the Hamiltonian components of $^5$He.
The difference between the results for the single AMD case shown in Table \ref{tab:TOAMD_E} and GCM is less than 2 MeV for each component.
Hence we can conclude that the GCM effect is not large in TOAMD because of the optimization of the correlation functions
with even a single AMD wave function.

\begin{table}[t]
  \begin{center}
    \caption{Hamiltonian components and radius of the ground and excited states of $^5$He in TOAMD+GCM
      in units of MeV and fm for energy and radius, respectively. The results of $^4$He are also shown for comparison.}
    \label{tab:GCM_H}
    \begin{tabular}{c|cccccc}
      \noalign{\hrule height 0.5pt}
      &~~Kinetic~~&~~Central~~&~~Tensor~~&~~Radius\\
      \noalign{\hrule height 0.5pt}
      ground state  &  $99.67$  & $-57.86$ & $-62.46$ &~~$2.11$       \\
      excited state &  $98.50$  & $-52.58$ & $-57.24$ &~~$2.43$       \\
      \noalign{\hrule height 0.5pt}
      $^4$He        &  $90.52$  & $-55.42$ & $-60.49$ &~~$1.54$       \\
      \noalign{\hrule height 0.5pt}
    \end{tabular}
  \end{center}
\end{table}

In the GCM calculation, we newly obtained the excited state of $^5$He, which has dominantly a positive parity component of 93\%.
The total energy, Hamiltonian components, and radius are listed in Tables~\ref{tab:GCM_E} and ~\ref{tab:GCM_H}.
The total energy is about $-11$ MeV, which is very close to the threshold energy consisting of $^3$H and $d$ of about $-10$ MeV, indicated by the arrow in Fig.~\ref{fig:ene_5He_E2}.
Experimentally, there is an excited $3/2^+$ state located just above the $^3$H+$d$ threshold energy by about 50 keV with a very small decay width of 75 keV \cite{tilley02}.
The obtained radius of the excited state is shown to be larger than that of the ground state by about 0.3 fm as shown in Table \ref{tab:GCM_H} under the bound state approximation. 
This could be an indication of the clustering state. 
We consider that the resulting excited state could be a candidate for the $^3$H+$d$ clustering state.
For the spin quantum number of the state, there might be a mixing of the $5/2^+$ state in the spectral function,
which is experimentally located above the $3/2^+$ state by about 2.3 MeV, although the $3/2^+$ state is regarded as a dominant component energetically.
We will investigate the detailed structure of this interesting state in a forthcoming paper.

We compare the Hamiltonian components between $^5$He and $^4$He to see the effect of a last neutron in Table~\ref{tab:GCM_H}.
The ground state of $^5$He shows larger values in magnitude for every Hamiltonian component, indicating more correlations than those of $^4$He by the interaction with a last neutron.
On the other hand, the excited state of $^5$He shows smaller interaction components than those of $^4$He.
This trend indicates the possibility of a different configuration from the $^4$He+$n$ one in the excited state.
It is important to note that the TOAMD framework can treat the ground and excited states in a unified manner
with a diagonalization of the Hamiltonian matrix under the variational principle.

\section{Summary}\label{sec:summary}
We have developed a new variational method of ``tensor-optimized antisymmetrized molecular dynamics'' (TOAMD) for nuclei,
which is a successive approach to treat bare nucleon--nucleon interactions without any renormalizations of the wave function and the interaction. 
Based on the successful results for $s$-shell nuclei, in this paper we have reported the first application of TOAMD to the $p$-shell nucleus, $^5$He, solving a five-body problem.
We prepare the reference AMD wave function of $^5$He with a $^4$He+$n$ cluster configuration and multiply the variational correlation functions of central and tensor types successively with up to the double products.

In TOAMD, the products of the Hamiltonian and the correlation function are expanded in a series of many-body operators using the cluster expansion.
We explicitly treat all of these operators in the calculation of the matrix elements in TOAMD
without any truncation of the higher-body operators.
This is an important point to keep the variational principle in the TOAMD wave function.
Owing to this advantage, we directly discuss the nuclear properties obtained in TOAMD without any transformation.

We further performed superposition of the TOAMD basis states with different $^4$He--$n$ distances of the AMD wave function.
It was found that effect of superposition on the total energy is less than 0.5 MeV.
This indicates that the important correlations are already described by the correlation functions.

We obtained not only the total energies but also the Hamiltonian components of the ground and excited states of $^5$He,
and discussed the effects of each correlation function on these quantities.
It is found that correlation functions always increase the contributions of central and tensor forces and the high-momentum components of $^5$He. 
We compared our calculations with those assuming common forms of the correlation functions similar to the Jastrow ansatz, which provides an energy loss of 4 MeV.
This means that independent optimization of the correlation functions is essential,
which can be general in the nucleon--nucleon correlations in nuclei.

It would be interesting to investigate the role of tensor correlation on the $LS$ splitting of $^5$He.
In our previous study with the tensor-optimized shell model \cite{myo05,myo11}, we discussed this effect with the bare AV8$^\prime$ potential;
the $3/2^-$ state gains a tensor force of more than about 4 MeV than the $1/2^-$ state, which results in a larger kinetic energy in $3/2^-$ than the $1/2^-$ state by about 7 MeV.
The obtained splitting energy including central and $LS$ components finally becomes 3 MeV.
In TOAMD, we will perform a similar analysis of the tensor correlation in relation to the clustering of $^5$He in the future. 

\section*{Acknowledgements}
We thank Professor Kiyomi Ikeda for fruitful discussions on this project. 
This work was supported by JSPS KAKENHI Grants No. JP18K03660.
Numerical calculations were performed partially on the computer system at RCNP, Osaka University.


%

\def\JL#1#2#3#4{ {{\rm #1}} \textbf{#2}, #3 (#4)}  
\nc{\PPNP}[3]   {\JL{Prog. Part. Nucl. Phys.}{#1}{#2}{#3}.}
\nc{\PTEP}[3]   {\JL{Prog. Theor. Exp. Phys.}{#1}{#2}{#3}.}

\end{document}